\documentclass[journal]{IEEEtran}

\usepackage{amssymb}
\usepackage{graphicx}
\usepackage{caption}
\usepackage{subcaption}
\usepackage{booktabs}% http://ctan.org/pkg/booktabs
\usepackage{multirow}
\usepackage[table,xcdraw]{xcolor}
\usepackage[normalem]{ulem}
\useunder{\uline}{\ul}{}
\newcommand\sbullet[1][.5]{\mathbin{\vcenter{\hbox{\scalebox{#1}{$\bullet$}}}}}

\begin{document}

%\title{Technological Trends and Key Communication Enablers for eVTOLs}
%\title{Key Technological Enablers for eVTOLs in Urban~Air~Mobility~(UAM)}
% \title{eVTOLs in Urban~Air~Mobility~(UAM): Requirements, Key Enablers, and Challenges}
% \title{eVTOLs in Urban~Air~Mobility: \\ Requirements, Key Enablers, and Challenges}

\title{eVTOL Communications and Networking in UAM:\\Requirements, Key Enablers, and Challenges}
\author{Abdullah~Abu~Zaid,
        Baha~Eddine~Youcef~Belmekki,
        and~Mohamed-Slim~Alouini
}

\maketitle
\thispagestyle{empty}
\begin{abstract}
Electric vertical takeoff and landing (eVTOL) aircraft have attracted great attention during the last years as the long-awaited enabler of urban air mobility (UAM), allowing a more sustainable method of transportation and accelerating the development of smart cities. The operation of eVTOLs over urban areas introduces safety hazards for passengers, pedestrians, and buildings, which prioritises safety considerations. Ensuring the safe operation of eVTOLs requires studying their communication, networking, computing requirements. In this paper, we showcase eVTOL requirements in UAM in terms of coverage, data rate, latency, spectrum efficiency, networking, and computing. Then, we identify potential key technological enablers to address these requirements and their challenges. Finally, we carry out a comparative case study between the terrestrial and aerial communication infrastructures to serve eVTOLs in UAM.

% The importance of safety in eVTOL operation cannot be understated, due to the operation of eVTOLs over urban areas creates safety hazards for passengers, pedestrians, and buildings.
% The importance of safety in eVTOL operation and the role of reliable communications are paramount [FILL]. 

% Still, while the excitement of taking to the skies grows and the development of eVTOLs accelerate, the significance of 
\end{abstract}

%\begin{IEEEkeywords}
%UAM, eVTOL, NTFP, UAV, HAP, LEO, RIS, NOMA, RSMA, SDN, NFV, RAN slicing, mmWave, THz, cloud computing, fog computing, MEC, digital twin.
%\end{IEEEkeywords}

\IEEEpeerreviewmaketitle

%\vspace{-0.5cm}
\section{Introduction}
The increase in global population, followed by urbanization and overcrowding, is steering the world toward sustainable smart city development. One enabler of such cities is urban air mobility (UAM). Several attempts have been made to realize air commuting, thus, circumventing congested urban areas and economizing on time and energy. However, limitations of conventional takeoff and landing (CTOL) aircraft, such as a need for a long runway, unsustainable and costly fuel, and loud noise, hindered their adoption for urban mobility. Nevertheless, the past decades witnessed technological advancements culminating in hundreds of companies designing and developing sustainable, electrically-powered vertical takeoff and landing (eVTOL) passenger aircraft. Predicted to enter large-scale service by the next decade, eVTOLs are thought to be the anticipated technology of future UAM.

UAM is defined as the use of low-altitude unmanned aerial vehicles (UAVs) and eVTOLs for cargo and passenger transportation in urban areas, complementing the current transportation systems to alleviate traffic congestion and reshape urban development sustainably \cite{NASAUAM2020reliable}. {UAM is a subset of advanced air mobility (AAM), which is a transformational vision covering all aspects of future unmanned vehicle operation in urban and rural areas.} The United States federal aviation authority (FAA) and national aeronautics and space administration (NASA) are coordinating to research and standardize all components of UAM. {This includes designing vertiports as eVTOL hubs and developing unique air traffic management (ATM) solutions for UAM's high-density and low-altitude operations.} UAM operations will utilize distributed traffic management by self-separating and self-navigating, aided by existing infrastructure.

{The UAM maturity level (UML) scale was developed by NASA to predict the evolution of UAM operation in terms of density, complexity, and automation. \cite{goodrich2021UML}} The density is described as the number of operating UAM aircraft in a given urban area. The operational complexity is a combination of number of vertiports, maximum vertiport capacity, weather tolerance level, and integration with other non-UAM vehicles. While automation degree defines the level of reliance on automated systems. Six levels were identified: UML-1 describes the certification and testing stage, while UML-2 to UML-6 correspond to increasing levels of density, complexity, and automation. For example, UML-2 predicts low-density and complexity operations with assistive automation, while UML-6 predicts ubiquitous operations with system-wide automation. We assume in this paper a UML-4 operation, which is characterized by medium density and complexity. In UML-4, eVTOLs will operate in air corridors, which are predefined aerial highways for UAM aircraft \cite{greenfeld2019concept}.

\begin{figure*}[ht!]
    \centering
    \includegraphics[width=\linewidth]{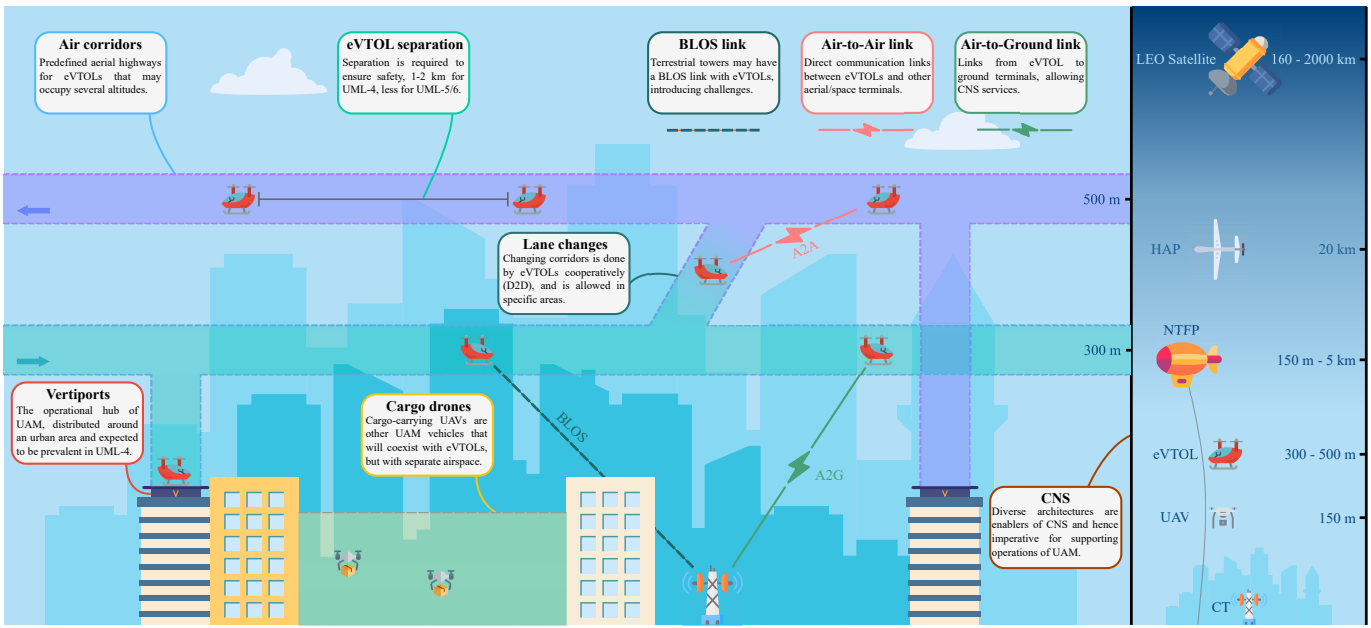}
    \caption{Illustration of a UAM vision for UML-4.}
    \label{fig:UAMillustration}
\end{figure*}

UAM will consist of numerous aircraft flying simultaneously in all directions. To ensure the safety of passengers and pedestrians alike, communication, networking, and computing solutions must be adapted to the unique characteristics of UAM and developed reliable air-to-air (A2A) and air-to-ground (A2G) transmission, ultimately enabling communication, navigation, and surveillance (CNS) services. Vertiports must also be equipped to assist eVTOLs in self-locating while approaching and landing. We aim in this paper to present communication, networking, and computing requirements in the context of eVTOLs in UAM, which are more stringent than other aircraft requirements in UAM, such as cargo drones, {and more demanding than general AAM requirements.} Hence, the studied requirements will offer an upper limit on the requirements of other UAM aircraft, {and typically a more flexible design of the system for AAM is possible.} We propose technological enablers for each requirement and list their advantages while taking into consideration their size, weight, and power (SWaP) limitations of onboard avionics. Furthermore, we address the challenges related to each key enabler. Finally, we conduct comparative simulations for three promising architectures: cellular towers (CTs), UAVs, and networked tethered flying platforms (NTFPs). An illustration of UAM is shown in Fig.~\ref{fig:UAMillustration} for a vision of future UML-4.

%\vspace{-0.3cm}
\section{eVTOL Requirements}
Reaching a ubiquitous and commercially viable UAM market will require innovations in eVTOL technology, regulatory standardization, urban infrastructure development, and increased public acceptance. Connecting eVTOLs in urban environments face different requirements and challenges than CTOL aircraft, helicopters, and UAVs. This is the first system that requires strict reliability, high rate, three-dimensional (3D) connectivity, low latency, and serving a high density of highly-mobile users simultaneously. Furthermore, from a hardware perspective, the primary objective while designing CNS avionics in eVTOLs is minimizing SWaP demands by ensuring that software updates are done over-the-air, without exchanging hardware boxes, and relying on ground-based technologies to aid in CNS services. We present several technological requirements in the context of eVTOLs.

%\vspace{-0.5cm}
\subsection{Coverage \& Reliability} The evolution of the aviation industry to become the safest transportation system gives an insight into how safety in UAM will be considered. At least the same strict regulations, standards, and safety records will be expected from eVTOL operations. The highly dynamic environment of UAM and the large number of obstacles in cities are key challenges in connecting eVTOLs. {A system offering continuous connectivity requires technologies that achieve beyond line-of-sight (BLOS) communications.} A reliable and wide coverage would ensure a continuous flow of necessary data for navigation and surveillance \cite{kim2022intelligent}, which is paramount for safe operations. The reliability requirement of UAM reaches $99.999\%$ \cite{baltaci2021survey}. We present several enablers of high coverage and reliability communications, such as next-generation cellular, aerial platforms, satellites, radio frequencies (RF), and reconfigurable intelligent surfaces (RIS).

%\vspace{-0.5cm}
\subsection{Data Rate} The applications of eVTOL communications will have varying data rate requirements. For example, voice applications and simple command and control (C2) messages can operate with low data rates. On the other hand, remote pilot operation (RPO) through video streaming and autonomous technology have more stringent requirements that reach $100$~Mbps \cite{baltaci2021survey}. Furthermore, passenger infotainment systems may be expected to provide high rates. The scarcity of bandwidth in cities is the primary challenge of enabling high data rates. Hence, we detail the different frequency ranges that will enable high throughput communications, such as millimeter-wave (mmWave) and terahertz (THz).
    
%\vspace{-0.5cm}
\subsection{Latency} The operation of eVTOLs over crowded cities and the low separation allowed in mature UAM, i.e. UML-5 and UML-6, will require much stricter latency demands than the aviation industry. Latency as low as $10$~ms is needed to enable RPO and autonomous technologies \cite{baltaci2021survey}. The challenge of achieving low-latency will be in reducing communication distances and eliminating relays. Decentralizing communications using sidelink technology is the first step toward achieving low-latency connectivity. Technologies studied in this paper that enable low latency are sidelink, and low-flying platforms such as UAVs and NTFPs.

%\vspace{-0.5cm}
\subsection{Spectrum Efficiency} The predicted future of high-density UAM and the internet of everything (IoE) signify a sharp increase of devices accessing the spectrum. Additionally, the capacity of the RF spectrum is reaching its limits, which only exacerbates the issue of spectrum scarcity. Current orthogonal multiple access schemes are not suitable to serve future eVTOL networks, {sparking a number of studies on the coexistence of terrestrial and aerial devices such as UAVs in 3D environments, which can easily be generalized for UAM.} Hence, this paper details two promising multiple access schemes: non-orthogonal multiple access (NOMA) and rate-splitting multiple access (RSMA).
    
%\vspace{-0.5cm}
\subsection{Networking} {At UML-4 and below, flight paths and vertiports are predetermined, making network design relatively simple. However, at UML-5 and above, the number of eVTOLs increases significantly, and at UML-6, operations become ad-hoc, or door-to-door, presenting significant network design challenges.} Consequently, more complex networking requirements have to be met. This pushes for a flexible and dynamic networking algorithm for UAM, which is realized by using software-defined networking (SDN) with network function virtualization (NFV). Other solutions detailed in the next section are sidelink and radio access network (RAN) slicing.
    
% %\vspace{-0.5cm}
\subsection{Computing} As technology progresses and more complex algorithms are developed, as well as the exponential increase in communication devices and network sizes, high computational power becomes crucial, especially for eVTOLs in urban areas. The supervision of UAM operation and offering navigation services to eVTOLs will be computationally-intensive. This prompted research into innovative computing paradigms for resource-limited devices in computations. Hence, we study technologies that enable eVTOLs to have access to strong computing power, such as cloud/fog computing (CC/FC), multi-access edge computing (MEC), and digital twins, without the need for expensive onboard hardware that defies SWaP requirements.

% \subsection{Hardware} The primary objective of CNS avionics design in eVTOLs is the minimization of SWaP demands.

\begin{table*}[ht!] 
\caption{eVTOL requirements and key enabling technologies.}
\centering
\label{tab:SummaryTable}
% \resizebox{\linewidth}{!}
{%
\begin{tabular}{llllcc}
\toprule
\multicolumn{1}{c}{\textbf{Requirement}} & \multicolumn{1}{c}{\textbf{Synergy with UAM}} & \multicolumn{1}{c}{\textbf{Key Technology}} & \multicolumn{1}{c}{\textbf{Advantage}} & \textbf{Maturity} & \textbf{Importance} \\ \toprule
 &  & \cellcolor[HTML]{EBEBEB}5G/6G & \cellcolor[HTML]{EBEBEB}Existing Large footprint of towers & \cellcolor[HTML]{EBEBEB}UML-2/UML-4 & \cellcolor[HTML]{EBEBEB}high \\
 &  & UAV/HAP & Flexible aerial coverage with low cost & UML-1 & high \\
 &  & \cellcolor[HTML]{EBEBEB}NTFP & \cellcolor[HTML]{EBEBEB}Secure aerial coverage with fiber backhaul & \cellcolor[HTML]{EBEBEB}UML-1 & \cellcolor[HTML]{EBEBEB}med \\
 &  & LEO & Expansive reliable coverage & UML-1 & high \\
 &  & \cellcolor[HTML]{EBEBEB}RIS & \cellcolor[HTML]{EBEBEB}Channel enhancer for BLOS operation & \cellcolor[HTML]{EBEBEB}UML-3 & \cellcolor[HTML]{EBEBEB}med \\
\multirow{-6}{*}{\begin{tabular}[c]{@{}l@{}}Coverage \&\\ Reliability\\ ($99.9$-$99.999\%$)\end{tabular}} & \multirow{-6}{*}{\begin{tabular}[c]{@{}l@{}}$\sbullet[.75]$ Wide connectivity\\ $\sbullet[.75]$ BLOS reachability\\ $\sbullet[.75]$ Safety\end{tabular}} & RF & Travels great distances and around obstacles & UML-1 & high \\ \hline
 &  & \cellcolor[HTML]{EBEBEB}mmWave & \cellcolor[HTML]{EBEBEB}Higher rate and security & \cellcolor[HTML]{EBEBEB}UML-2 & \cellcolor[HTML]{EBEBEB}high \\
\multirow{-2}{*}{\begin{tabular}[c]{@{}l@{}}Data Rate\\ ($<100$ Mbps)\end{tabular}} & \multirow{-2}{*}{\begin{tabular}[c]{@{}l@{}}$\sbullet[.75]$ Video streaming\\ $\sbullet[.75]$ Entertainment\end{tabular}} & THz & Near infinite bandwidth & UML-4 & low \\ \hline
 &  & \cellcolor[HTML]{EBEBEB}Sidelink & \cellcolor[HTML]{EBEBEB}Direct D2D link achieves lowest latency & \cellcolor[HTML]{EBEBEB}UML-2 & \cellcolor[HTML]{EBEBEB}high \\
 &  & UAV & Low altitude and close to eVTOLs & UML-1 & med \\
\multirow{-3}{*}{\begin{tabular}[c]{@{}l@{}}Latency\\ ($10$-$500$ ms)\end{tabular}} & \multirow{-3}{*}{\begin{tabular}[c]{@{}l@{}}$\sbullet[.75]$ RPO\\ $\sbullet[.75]$ Safety\\ $\sbullet[.75]$ Autonomy\end{tabular}} & \cellcolor[HTML]{EBEBEB}NTFP & \cellcolor[HTML]{EBEBEB}{High-speed tether link} & \cellcolor[HTML]{EBEBEB}UML-1 & \cellcolor[HTML]{EBEBEB}med \\ \hline
 &  & NOMA & Unlock New dimension of spectrum access & UML-1 & med \\
\multirow{-2}{*}{\begin{tabular}[c]{@{}l@{}}Spectrum \\ Efficiency\end{tabular}} & \multirow{-2}{*}{\begin{tabular}[c]{@{}l@{}}$\sbullet[.75]$ High user density\\ $\sbullet[.75]$ Cost effective\end{tabular}} & \cellcolor[HTML]{EBEBEB}RSMA & \cellcolor[HTML]{EBEBEB}Flexible spectrum access algorithm & \cellcolor[HTML]{EBEBEB}UML-1 & \cellcolor[HTML]{EBEBEB}med \\ \hline
 &  & SDN/NFV & High flexibility and programmability & UML-1 & high \\
 &  & \cellcolor[HTML]{EBEBEB}Sidelink & \cellcolor[HTML]{EBEBEB}Network coverage increase via multi-hop & \cellcolor[HTML]{EBEBEB}UML-2 & \cellcolor[HTML]{EBEBEB}high \\
\multirow{-3}{*}{Networking} & \multirow{-3}{*}{\begin{tabular}[c]{@{}l@{}}$\sbullet[.75]$ Decentralization\\ $\sbullet[.75]$ Autonomy by D2D\end{tabular}} & RAN slicing & Virtual independent network requirements & UML-2 & med \\ \hline
 &  & \cellcolor[HTML]{EBEBEB}CC \& FC & \cellcolor[HTML]{EBEBEB}High computing power complying with SWaP & \cellcolor[HTML]{EBEBEB}UML-1 & \cellcolor[HTML]{EBEBEB}low \\
 &  & MEC & Low-latency computing power & UML-1 & med \\
\multirow{-3}{*}{Computing} & \multirow{-3}{*}{\begin{tabular}[c]{@{}l@{}}$\sbullet[.75]$ Weather prediction\\ $\sbullet[.75]$ Route planning\end{tabular}} & \cellcolor[HTML]{EBEBEB}Digital twin & \cellcolor[HTML]{EBEBEB}Overarching supervision of UAM system & \cellcolor[HTML]{EBEBEB}UML-4 & \cellcolor[HTML]{EBEBEB}low \\ \bottomrule
\end{tabular}%
}
\end{table*}

\section{Key Enabling Technologies}
This section presents the key enablers sectioned into their respective categories: architecture, air interface, networking, spectrum, and computing paradigm. Table~\ref{tab:SummaryTable} summarizes the key technologies and provides pertinent information such as their advantage, phasing timeline, and importance to eVTOLs.  

%\vspace{-0.5cm}
\subsection{Architecture}

\subsubsection{Cellular Towers}
The large-scale footprint of cellular towers (CTs) makes them a prominent solution for wide coverage eVTOL communications. CTs are widely deployed in urban areas, and can provide eVTOLs with broad connectivity across a city, especially around vertiports, to enable CNS services. Current cellular networks such as fifth-generation (5G) boast increased data rates, low latency, and support for higher mobility. Thus, it is recommended for eVTOL communications, albeit with modifications to antenna and priority setups \cite{NASAUAM2020reliable}. The next generation of cellular networks, i.e., sixth-generation (6G), is envisioned to enable 3D connectivity for aerial platforms, ultra-high density support that enables operation of UML-5 and beyond, and extremely low latency for reliable C2 and autonomous operation. Researchers are targeting the start of the next decade for maturity and deployment of 6G, which aligns with the prediction of the industry to start large-scale operations of UAM in the same period, thus, allowing UAM to utilize 6G.

\noindent
\textbf{Challenges}: Existing CTs are unsuitable for aerial connectivity since antennas are tilted downwards for terrestrial users \cite{NASAUAM2020reliable}. Either existing towers must be reconfigured to serve aerial vehicles, or new dedicated towers should be built, preferably on rooftops. Furthermore, aviation standards stipulate that aerial communications must be prioritised, especially in emergencies, which is not supported by current cellular networks. Finally, having numerous communicating nodes at high altitudes introduces strong interference on ground users, due to the high probability of LOS.

\subsubsection{Unmanned Aerial Vehicles \& High Altitude Platforms}
UAVs are airborne, small-size, low-cost vehicles usually deployed in swarms and distributed over an area to maximize coverage. They operate at altitudes around $150$~m to minimize latency and battery usage. Free-flying UAVs can be dynamically reassigned to different areas. Furthermore, The closeness of operational altitudes of UAVs and eVTOLs allows for low-latency line-of-sight (LOS) links  to be used for communication and surveillance of UAM aircraft \cite{kim2018designing}. On the other hand, high altitude platforms (HAPs) operate in the stratosphere at an altitude of $20$~km. Their high elevation angles allow for a larger coverage area than UAVs and a more reliable LOS link. HAPs may operate as relays that connect satellite systems with eVTOLs. The flexibility and low operational cost of UAVs and HAPs make them an attractive solution with reliable communications and navigation technologies for eVTOLs.

\noindent
\textbf{Challenges}: {Densely operating eVTOLs alongside UAVs in urban cities poses safety concerns due to the risk of hijacking and collisions. Additionally, limited flying time reduces UAV availability and reliability, while the introduction of HAPs may increase delay and become expensive at scale.}

% Deploying hundreds or thousands of low-altitude UAVs in urban cities presents a safety issue for a dense operation of eVTOLs. UAVs may be hijacked by wrongdoers which can cause collisions with eVTOLs. Another issue is the limited flying time of UAVs, reducing their availability and reliability. HAPs introduce more delay than UAVs, and large-scale deployment may become expensive.

\subsubsection{Networked Tethered Flying Platforms}
NTFPs are flying platforms connected to the ground via a tether that provides continuous power and data. This will increase flying time, backhaul capacity, and security compared to free-flying platforms \cite{belmekki2022unleashing}. Furthermore, NTFPs offer higher endurance which allows persistent communication services for eVTOLs, increasing reliability. NTFPs come in several types: tethered UAVs, tethered balloons, tethered blimps, and tethered Helikites. Depending on the application, NTFPs may operate at altitudes ranging from $150$~m to $5$~km. The dedicated backhaul link in the tether reduces communication latency and enhances performance by eliminating interference from other sources.

\noindent
\textbf{Challenges}: The tether is a notable limitation for NTFPs, preventing flexible and dynamic deployment. Furthermore, the tether may become a hazard for flying objects such as eVTOLs, prohibiting their deployment over UAM air corridors.

\subsubsection{Low Earth Orbit Satellites}
Low Earth orbit (LEO) satellites orbit relatively close to Earth, from around $160$~km up to $2000$~km. This high altitude facilitates coverage for inter-city or suburban eVTOL trips. A large number of LEO satellites are typically deployed to make up a constellation offering near-continuous connectivity and navigation. The advantage of using satellites over aerial platforms is the increased service time, which is typically $15$~years. LEO satellites may use spot beam technologies that enable high data rates while having latency similar in magnitude to traditional ground networks.

\noindent
\textbf{Challenges}: {Low altitude satellite communication is expensive and requires a large constellation, leading to issues such as strong Doppler effects, fading, and rapid handovers. In addition, special hardware is required to connect to satellites, which may not meet eVTOL SWaP requirements.}

%\vspace{-0.4cm}
\subsection{Air Interface}

\subsubsection{Non-Orthogonal Multiple Access}
NOMA allows users to access the channel simultaneously using the same frequency and time resources, however, with different power levels \cite{saito2013system}. By utilizing the same resources, NOMA achieves higher spectral efficiency than traditional orthogonal access schemes. Furthermore, lower latency is achieved since users do not need to wait for their time slots to transmit information. NOMA may be used according to two priority rankings: channel ranking and quality of service (QoS) ranking. When applying the channel ranking, eVTOLs with weaker channel gains will transmit with a higher power and vice-versa to achieve fairness. For the QoS ranking, eVTOLs will be ranked according to their needs in terms of QoS. For instance, an eVTOL requiring urgent transmission will be assigned more power than an eVTOL with less urgent transmission.

\noindent
\textbf{Challenges}: Receiver complexity is a notable disadvantage of NOMA since a user must decode other users' messages to decode their own, increasing delay and power usage. For high-density UAM, NOMA will suffer from increased complexity and delay, especially for eVTOLs with low QoS ranking, making NOMA unsuitable for C2 links.

\subsubsection{Rate-Splitting Multiple Access}
RSMA is a multiple access technique similar to NOMA, where users can send simultaneously on the same resources. However, in RSMA, receivers may decode part of the interference and treat the other part as noise \cite{mao2022rate}. RSMA is a scheme in-between traditional multiple access schemes and NOMA, where traditional schemes fully treat the interference as noise, and NOMA fully decodes the interference. The flexibility of RSMA enables good performance for all magnitudes of interference. Hence, eVTOLs may utilize RSMA by adjusting the decoding level for the different parts of the trip since urban areas usually contain higher interference than suburban areas.

\noindent
\textbf{Challenges}: {RSMA receivers are more complex than NOMA receivers, and the complexity increases with the number of users. Additionally, RSMA has high encoding complexity due to message splitting, which requires extra signaling overhead} \cite{mao2022rate}. Hence, RSMA-aided eVTOLs may be restricted by SWaP requirements.

\subsubsection{Reconfigurable Intelligent Surfaces}
Advancements in wireless communications were primarily applied at the transmitter and the receiver while the channel was untouched. Recently, RIS gained popularity as a way to alter the channel to enhance the performance of wireless communications \cite{wu2019towards}. RIS is a low-cost passive reflective surface that works by changing the signal's phase or amplitude to amplify the signal in the receiver's direction. RIS technology is a promising enabler for improving coverage in BLOS communications in urban environments, where buildings often block LOS. The reconfigurable nature of RIS allows for live-tracking of eVTOLs to maintain the communication link while cruising. Furthermore, RIS can be mounted on the side of buildings, rooftops, billboards, and UAVs to increase coverage.

\noindent
\textbf{Challenges}: RIS need channel state information to operate, but since eVTOLs travel at high speeds in 3D, large Doppler shifts will hinder their operation. RIS also introduces propagation and processing delays unsuitable for C2 communications.

%\vspace{-0.3cm}
\subsection{Networking}

\subsubsection{Sidelink}
One of the most crucial technologies that will assist eVTOL communications and navigation is sidelink, also termed, device-to-device (D2D). By decentralizing the network, D2D networks allow eVTOLs to communicate directly with each other, minimizing delay and increasing coverage via multi-hop. Connecting to core networks such as cellular might not always be possible for eVTOLs, leaving D2D as the only option. Furthermore, D2D provides a higher rate and lower latency when eVTOLs are in close proximity; this is especially important for autonomous applications, such as collision avoidance and eVTOL platooning in air corridors.

\noindent
\textbf{Challenges}: In a D2D eVTOL network, interference management, requirements, and performance will change adversely with an increasing number of eVTOLs. A dense UAM operation such as UML-5 and above will suffer delays due to extensive resource sharing. Furthermore, SWaP constraints of eVTOLs will limit the transmission range. 

\subsubsection{Software-Defined Networking \& Network Function Virtualization}
Achieving desired qualities of eVTOL networks, such as flexibility and programmability, is difficult since traditional network functionalities are implemented using dedicated hardware that requires manual configuration. SDN uses software to control and intelligently reconfigure the network. This is done by separating the data plane from the control plane. SDN enables programmable, scalable, and easily upgradable networks for eVTOLs. Alongside SDN, NFV is a network architecture that offloads software functions to virtual machines in servers, thus, separating network services from hardware and lowering the cost of operation \cite{zhuang2019sdn}. Hence, NFV allows easy implementation of new technologies without needing new hardware. NFV is a promising network architecture for eVTOL communications, where new functionalities can be integrated using software, allowing scalable and cost-effective eVTOL networks.

\noindent
\textbf{Challenges}: {Virtualization faces challenges in complexity, latency, and security.} Other issues stem from the susceptibility of the SDN controller to failure, which would be detrimental to eVTOLs, and the need for standardization of SDN across all UAM aircraft.

\subsubsection{Radio Access Network Slicing}
RAN slicing utilizes SDN/NFV technologies to segment the network infrastructure into multiple virtual end-to-end networks, each with different use cases. Each slice of the network can be assumed as a separate network, which enables different applications with differing connectivity requirements to function on the same network hardware. This is ideal for eVTOL applications since they differ vastly in terms of the required rate, reliability, security, and latency. {For instance, safety information for control towers requires low-latency, passenger infotainment requires high data rate, and location sharing between eVTOLs requires low-latency and high-reliability.} To address this, the network resources can be distributed appropriately to each application. Additionally, RAN slicing improves security by isolating the attacks to a single slice and leaving the other slices secure.

\noindent
\textbf{Challenges}: {Further research is needed to implement RAN slicing in practice and address associated challenges.} Notably, increasing the number of slices will increase the management complexity. Furthermore, a mobility-aware network slicing design is crucial for eVTOLs.

\begin{figure*}[ht!]
     \centering
     \begin{subfigure}[b]{0.35\textwidth}
         \centering
         \includegraphics[width=1\textwidth]{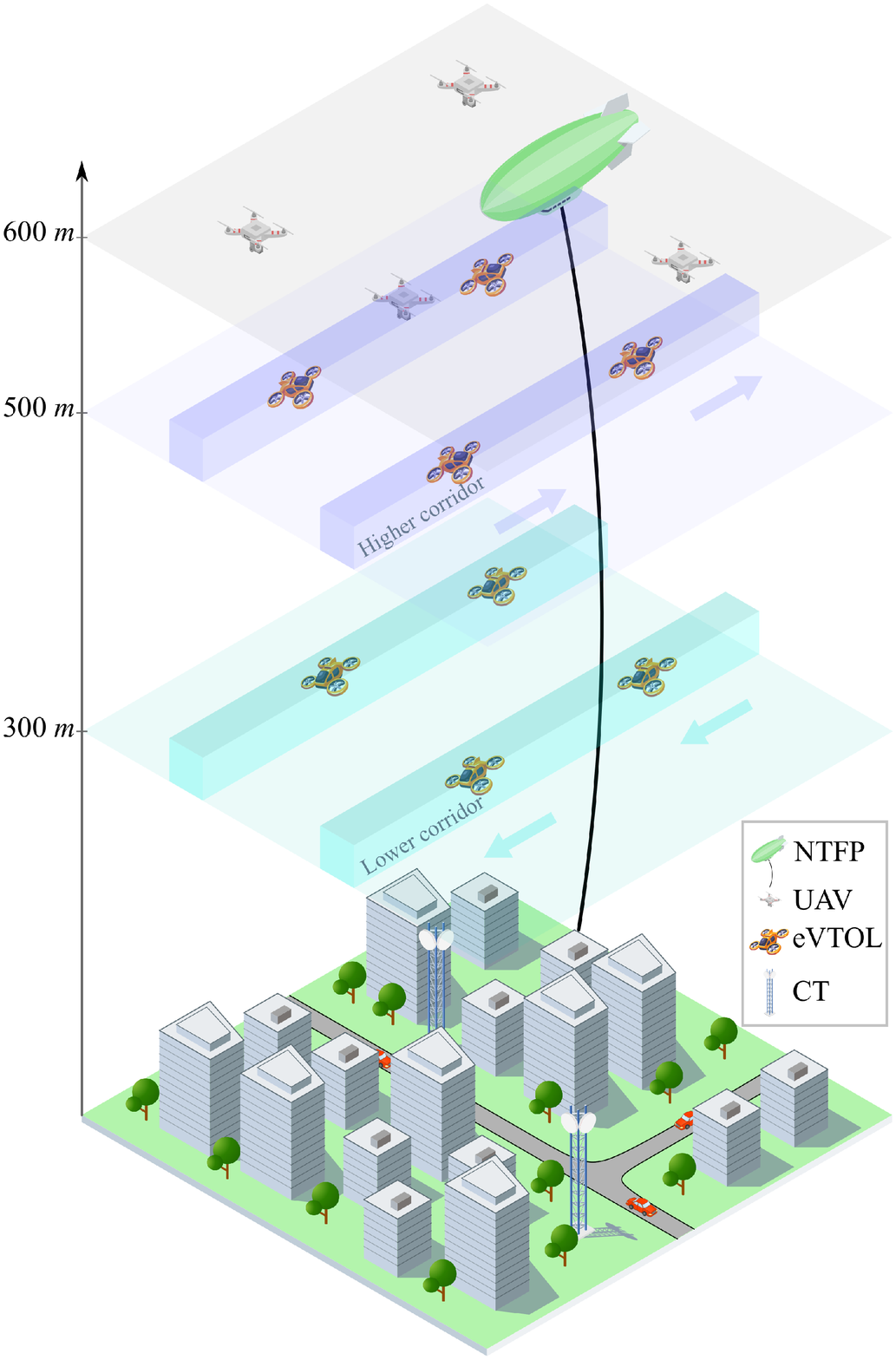}
         \caption{Illustration of the considered system model.}
         \label{fig:SimFig1}
     \end{subfigure}
     % \hfill
     \begin{subfigure}[b]{0.42\textwidth}
         \centering
         \includegraphics[width=1\textwidth]{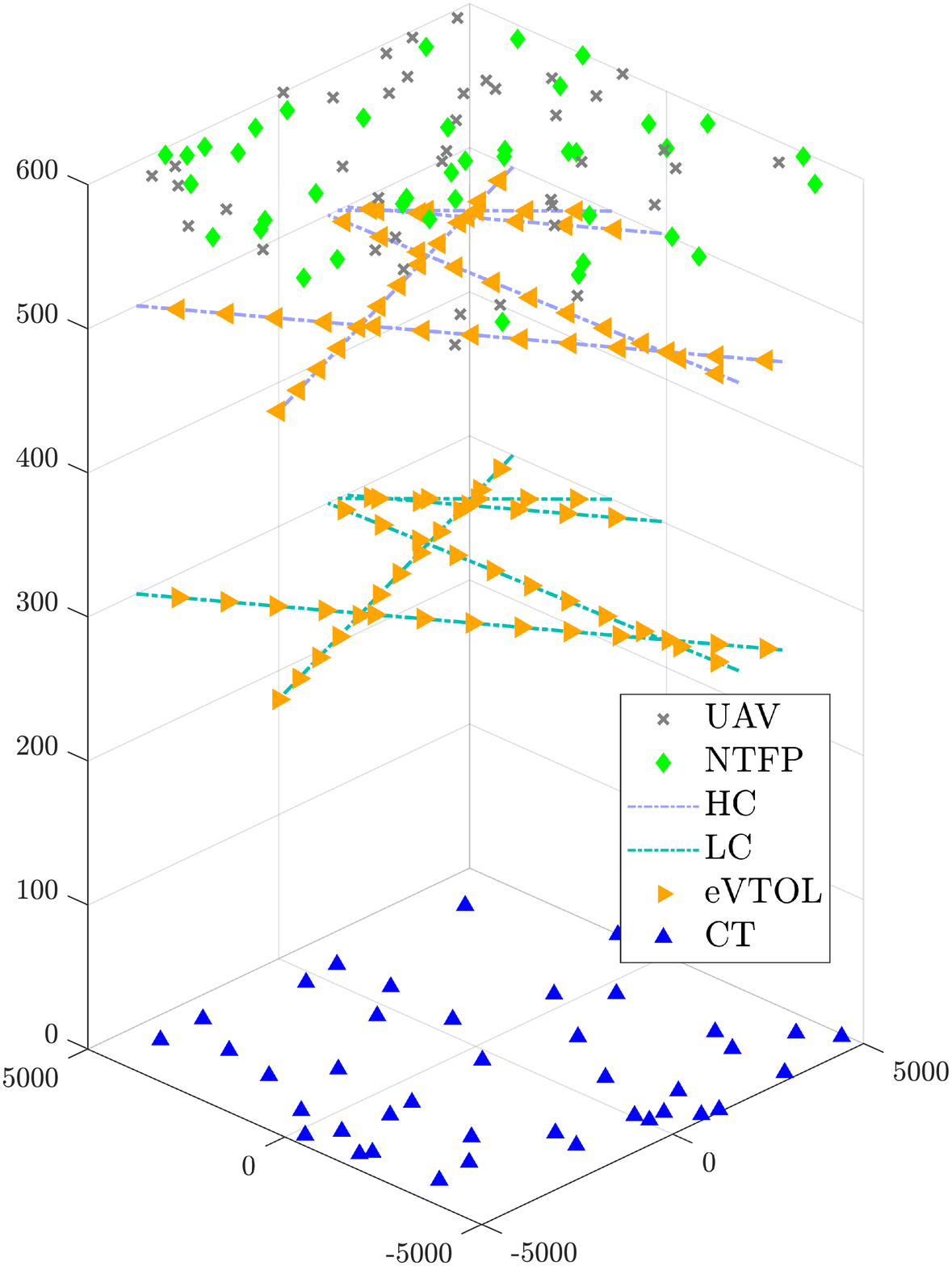}
         \caption{Simulation abstraction of the system model.}
         \label{fig:SimFig2}
     \end{subfigure}
        \caption{A Potential architecture for eVTOLs in UAM using CTs, UAVs, and NTFPs.}
        \label{fig:SimulationFigures}
\end{figure*}

%\vspace{-0.3cm}
\subsection{Spectrum}

\subsubsection{Radio Frequency}
Technologies utilizing RF frequencies, which range from $20$~kHz to $30$~GHz, are an integral part of communications in UAM. RF signals offer large coverage by traveling great distances, penetrating walls, and reaching behind obstacles, enabling BLOS operations. RF is extensively used in the aviation industry; for example, automatic dependent surveillance-broadcast (ADS-B), which operates on $1090$~MHz, enables cooperative vehicle-to-vehicle separation. However, for ADS-B to be used effectively in dense UAM, modifications must be made to alleviate spectrum congestion. Specifically, since the aim of ADS-B is cooperative separation of other neighboring aircraft, eVTOLs can significantly reduce the transmitted power relative to traditional aviation standards \cite{NASAUAM2020reliable}. Another technology that can support separation and surveillance applications of UAM is the second universal access transceiver (UAT2), reserved on $1104$~MHz. Furthermore, very high frequency (VHF) data link (VDL) is used for aeronautical communication and is reserved by the FAA from $118$~MHz to $137$~MHz. VDL mode 2 is most widely used by industry and supports digital data transfer for A2G links. VDL mode 3 was developed with higher spectral efficiency and supported digital voice and data; however, it was not implemented due to a lack of interest from the aviation industry. On the higher end of RF, the C-band from $5030$~MHz to $5091$~MHz is studied by the FAA for C2 communications. 

\noindent
\textbf{Challenges}: Current RF technologies, such as ADS-B, cannot accommodate the densities of UML-4 and will be overwhelmed with the dramatic increase of UAM aircraft. Furthermore, some frequencies, such as VDL mode 3, are unutilized and must be reallocated for practical use in UAM \cite{NASAUAM2020reliable}.

\subsubsection{Millimeter-Wave}
% mmWave uses the frequency range from around $30$~GHz to $300$~GHz, offering a much larger bandwidth than RF and offering potentially very high data rates. Furthermore, mmWave signals have very short wavelengths, making small antennas possible, achieving SWaP requirements on eVTOLs and allowing several antennas to improve capacity and range.
{mmWave, with frequencies between 30 GHz and 300 GHz, provides greater bandwidth than RF and high data rates. Its short wavelengths enable small antennas, meeting eVTOL SWaP requirements and enabling multiple antennas to improve range and capacity.} An emerging technology currently being tested is frequency-modulated continuous wave (FMCW) mmWave radar, a low-cost, small-footprint radar used for tracking and surveilling other aircraft, flying objects, and obstacles. The high frequency of FMCW radar enables the detection of multiple objects in highly dense situations \cite{NASAUAM2020reliable}. However, due to the strong attenuation of high frequencies, highly directional beamforming techniques must be implemented \cite{al2016millimeter}. Fortunately, this allows for highly secure links and makes intercepting communication difficult. Additionally, the large bandwidths available for mmWave have the capacity to serve highly dense environments, such as UAM.

\noindent
\textbf{Challenges}: Further research is needed to achieve the high potential of mmWave technologies. Challenges that need to be addressed, which are exacerbated in eVTOLs, are beam misalignment due to device wobbling, high power consumption, and sensitivity to blockage and vibrations.

\subsubsection{Terahertz}
Defined roughly as frequencies between $100$~GHz and $10$~THz and containing abundant bandwidths, THz frequencies present an opportunity for extremely high data rates {\cite{lou2023coverage}}. However, due to the severe attenuation characteristics and strict reliance on LOS, THz is mainly envisioned for fixed point-to-point links. Hence they may be implemented at vertiports when eVTOLs are stationary for high-throughput data offloading. THz may operate in a complementary fashion in eVTOL networks by enabling high-throughput wireless backhaul links for HAPs and satellites. Furthermore, THz with large antenna arrays enables highly precise localization, improving navigation services for eVTOLs.

\noindent
\textbf{Challenges}: 
{THz communication is in the early stages of research and deployment, limited by high path loss, range and inability to penetrate objects. Additionally, communication circuits consume more power at higher frequencies.}

%\vspace{-0.3cm}
\subsection{Computing Paradigm}

\subsubsection{Cloud Computing \& Fog Computing}
Cloud computing offers computational resources to devices, often deployed on the internet. The devices send their data to the cloud, where the required computations are performed, and results are disseminated back if needed. The safe operation of a UAM system will involve computationally-intensive algorithms and operations, such as end-to-end route planning and weather analysis and prediction. eVTOLs can offload such delay-tolerant algorithms to the cloud. Similarly, fog computing provides computational resources closer to the end devices, reducing latency and bandwidth usage on the network. A trade-off exists between using cloud computing over fog computing, where the latter has less computational capabilities but faster response time.

\noindent
\textbf{Challenges}: Typically, high-performance computing devices are situated far from end users, increasing the delay and system bandwidth usage. This makes cloud services unable to serve delay-intolerant applications. Furthermore, the high mobility of eVTOLs poses a challenge for the UAM system to locate the eVTOL and disseminate back the computations.

\subsubsection{Multi-access Edge Computing}
Rather than off-loading computations to a center far away and introducing delay, MEC technology has distributed computing servers at the edge of the network, in close proximity to communication devices. {MEC servers are typically located at cellular base stations (BSs), but there have been recent proposals to install them on UAVs.} \cite{zhou2020mobile}. To avoid excessive handovers at high speeds, MEC technology is most suitably used at vertiports when eVTOLs are taking off or landing. Additionally, the expansive coverage of HAPs may allow serving eVTOLs with computing power without the complexity of handovers.

\noindent
\textbf{Challenges}: Pushing computing power close to user devices requires considerable investment since a multitude of MEC nodes must be deployed to serve a highly dense UAM system. Additionally, the mobility of eVTOLs restricts the use of MEC technology while cruising.

\subsubsection{Digital Twin}
A digital twin is a virtual real-time copy of a physical object that can be used to track and synchronize with the original object \cite{wang2020digital}. In a multi-device network, devices send their data to the digital twin which builds a virtual world where all devices are synchronized. A digital twin for eVTOLs would run simulations of the UAM system and provide the participants with beneficial data, such as optimal route planning, possible congestion situations, and collision predictions. The digital twin can also be used to predict the performance of the UAM communication network, and optimize it accordingly.

\noindent
\textbf{Challenges}: {Digital twins are not yet feasible at large scales due to the large data requirements, intensive computations of a dense UAM, and security concerns as it contains information about all users and access by an ill-intentioned entity is a considerable security risk.}

%\vspace{-0.3cm}
% \section{Case Study: Terrestrial and  Aerial Coverage for eVTOLs}
\section{Case Study: Terrestrial and  Aerial Coverage}
In this section, we provide a comparative simulation for using CTs, UAVs, and NTFPs as BSs for eVTOLs in a dense urban environment that supports the operation of UML-4. We assume all users operate on RF frequency and are equipped with UAT2 transceivers, while BSs maintain a high-throughput, low-latency backhaul link. We assume CT antennas have beams directed upwards. We use MATLAB to simulate the case study. To this end, we assess the downlink (DL) and uplink (UL) performance using the coverage probability metric, which measures the probability of a user to achieve a certain signal-to-interference ratio (SIR) threshold. We use the Poisson line process (PLP) to model the location of eVTOLs, where each line corresponds to an air corridor with two levels at $300$~m and $500$~m, corresponding to two ways of traffic. All BSs are distributed according to a two-dimensional (2D) Poisson point process (PPP). However, any NTFP within $200$~m horizontally from an air corridor is removed due to the safety hazard of the tether. CTs are at a height of $40$~m while UAVs and NTFPs are at a height of $600$~m. The density of BSs is set to $40~\textrm{BS}/100~\textrm{km}^2$. Fig.~\ref{fig:SimulationFigures} illustrates a realistic system model used in the simulation. Two channel models are used depending on the existence of a LOS link or a non-line-of-sight (NLOS) link. In LOS link, we use Nakagami-$m$ to model small-scale fading and a path loss exponent of $2$, whereas in NLOS link, we use a Rayleigh model with a path loss exponent of $4$. An elevation-based model using a Sigmoid function is used to model the probability of LOS between CTs and eVTOLs \cite{al2014optimal}, while links between aircraft are considered in LOS due to their high altitudes and the scarcity of obstacles.

\begin{figure}[t!]
    \centering
    \includegraphics[width=1\linewidth]{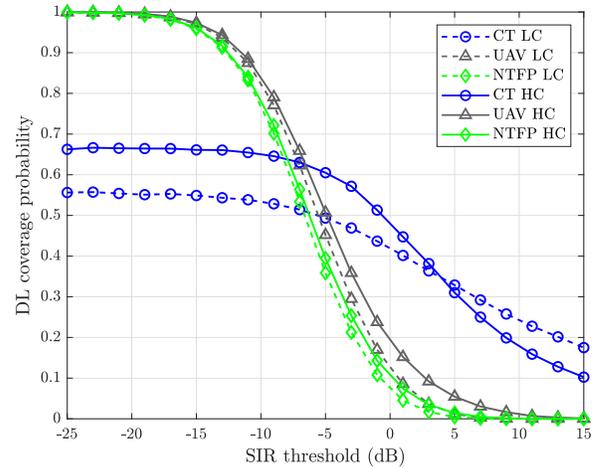}
    \caption{Coverage probability for the DL transmission.}
    \label{fig:simulationfig3}
\end{figure}

\begin{figure}[h!]
    \centering
    \includegraphics[width=1\linewidth]{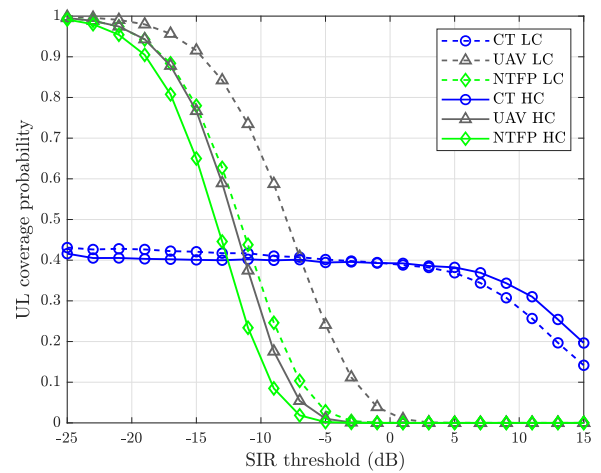}
    \caption{Coverage probability for the UL transmission.}
    \label{fig:simulationfig4}
\end{figure}

In Fig.~\ref{fig:simulationfig3} and Fig.~\ref{fig:simulationfig4}, we show the coverage probabilities of DL and UL, respectively, and we plot for each BS performance a low-corridor (LC) and a high-corridor (HC) curve. A general trend seen in both figures signifies the advantage of using NTFPs and UAVs (aerial BSs) over CTs for low SIR. For high SIR, however, strong interfering aerial devices with LOS links hinder the performance of UAVs and NTFPs, making CTs' coverage more resilient in this case. Furthermore, we see an inverse of performance between LC and HC eVTOLs for DL and UL. HC eVTOLs achieve better performance than LC eVTOLs for DL and vice-versa for UL. This is due to the change in distance between devices in the corridors. Lastly, we notice the drop in performance of NTFPs compared to UAVs due to the inability of NTFPs to operate directly over air corridors. In conclusion, it is advantageous for aerial and terrestrial BSs to be used synergistically in a UAM system to improve coverage in low and high SIR levels.

Although the above simulation was carried out in MATLAB, there are other simulators and testbeds available that can accelerate the development of UAM and aid in understanding the behavior and requirements of eVTOLs. Several platforms have been developed. For instance, the national science foundation (NSF) launched the aerial experimentation and research platform for advanced wireless (AERPAW), a 5G-supported testbed to study UAV communications and ATM. Another project is 5G!Drones, which allows studying different UAV use-cases over 5G while utilizing network slicing technology.

%\vspace{-0.4cm}

\section{Conclusion}
This paper discussed eVTOL vehicles in UAM and their role in future sustainable smart cities. Furthermore, their communication, networking, and computing requirements were presented. Then, key enabling technologies were identified, and their significance for eVTOLs was demonstrated. Specifically, CTs, UAVs/HAPs, NTFPs, and LEO satellites were presented as architectural solutions. NOMA, RSMA, and RIS were discussed as air interfaces. Sidelink, SDN/NFV, and RAN slicing were recognized as networking solutions. Explored spectrum frequencies were RF, mmWave, and THz, while computing demands were addressed with cloud/fog computing, MEC, and digital twins. In addition, challenges specific to each key technology were studied. Finally, using simulations, a case study was carried out to compare the use of terrestrial and aerial BSs for eVTOLs. It showed that each BS type is advantageous in specific cases; thus, a combination of BSs might yield the best performance.

%\ifCLASSOPTIONcaptionsoff
  %\newpage
%\fi

%\vspace{-0.3cm}
\bibliographystyle{IEEEtran}
\bibliography{refs_Abbrev.bib}
%\vspace{-0.5cm}
\section*{biographies}
%\vspace{-1.2cm}
\begin{IEEEbiographynophoto}
{Abdullah Abu Zaid}
(abdullah.abuzaid@kaust.edu.sa) is a Ph.D. student in the Communication Theory Lab (CTL) under the supervision of Professor Mohamed-Slim Alouini at King Abdullah University of Science and Technology (KAUST). Abdullah obtained his master's degree in Electrical and Computer Engineering from KAUST in 2021 and his bachelor's degree from The University of Jordan in 2020. During his engineering studies, he did research in wireless sensor networks and cognitive radio. His current research interests include stochastic geometry modeling, and flying platforms.
\end{IEEEbiographynophoto}
%\vspace{-1.2cm}
\begin{IEEEbiographynophoto}
{Baha Eddine Youcef Belmekki}
(bahaeddine.belmekki@kaust.edu.sa) is currently a postdoctoral research fellow at Communication Theory Laboratory of KAUST. His research interests include cooperative communications, millimeter-wave communications, non-orthogonal multiple access systems, stochastic geometry analysis of vehicular networks, and networked flying platforms.
\end{IEEEbiographynophoto}
%\vspace{-1cm}
\begin{IEEEbiographynophoto}
{Mohamed-Slim Alouini} 
(slim.alouini@kaust.edu.sa) received his Ph.D. degree in electrical engineering from the California Institute of Technology, Pasadena, in 1998. He is now a Distinguished Professor of Electrical Engineering at KAUST. His current research interests include the modeling, design, and performance analysis of wireless communication systems.
\end{IEEEbiographynophoto}

\end{document}